\documentclass[letterpaper, 10 pt, conference]{ieeeconf} 

\IEEEoverridecommandlockouts                              

\usepackage{amsmath,amssymb,amsfonts}
\usepackage{algorithm}
\usepackage{cuted}
\usepackage{capt-of}
\usepackage{algpseudocode}
\usepackage{booktabs} 
\usepackage{svg}
\usepackage{adjustbox}
\usepackage{array}
\usepackage{balance}
\usepackage{graphicx}
\usepackage{tikz}
\usepackage{pgfplots}
\pgfplotsset{compat=1.17}
\usepackage{hyperref}
\usepackage{authblk}
\usepackage{cite}
\usepackage{comment}
\usepackage{cleveref}
\usepackage{dirtytalk}
\usepackage{dblfloatfix}

\title{\LARGE \bf \adjustbox{max width=\textwidth}{Co-Designing Social Robots for Social-Cognition Training with Autistic Adults}}

\begin{document}

\author{
\begin{minipage}{0.88\textwidth}
\centering
Yuval Zohar$^{1}$ \quad
Mordi Benhamou$^{2}$ \quad
Guy Laban$^{1,2,3,*}$\\

\footnotesize
$^{1}$Department of Industrial Engineering and Management, Ben-Gurion University of the Negev, Beer Sheva, Israel\\
$^{2}$The Azrieli National Center for Autism and Neurodevelopment Research, Ben-Gurion University of the Negev, Beer Sheva, Israel\\
$^{3}$School of Brain Sciences and Cognition, Ben-Gurion University of the Negev, Beer Sheva, Israel\\

\footnotesize
$^{*}$Corresponding author: \href{mailto:laban@bgu.ac.il}{\texttt{laban@bgu.ac.il}}
\end{minipage}
}

\maketitle

\thispagestyle{empty}
\pagestyle{empty}

\begin{abstract}
Social robots have been widely explored as tools for autism intervention, yet this literature has focused predominantly on children and has rarely involved autistic adults as active contributors to design. This creates a mismatch between existing systems and the social-cognitive challenges autistic adults actually face in everyday life, including navigating ambiguous interpersonal contexts, managing conversational timing, and interpreting implied emotional meaning. To address this gap, we conducted an online focus group and co-design session with five autistic adults to explore what a social robot for social-cognition training should do, how it should interact, and under what conditions it would be genuinely useful. The 90-minute session combined open discussion with structured co-design activities on a shared digital whiteboard, and the resulting verbal and visual data were analysed using reflexive thematic analysis. The analysis yielded seven themes that define core design requirements: the robot should function as a scaffold rather than a substitute, prioritise authenticity over comfort, provide personalised and user-controlled feedback, accommodate emotional self-awareness gaps, respect privacy and contextual boundaries, support rehearsal for real-world social situations, and remain configurable in identity, form, and expression. Together, the findings suggest that autistic adults envision the robot not as a companion or live social assistant, but as a private, configurable rehearsal partner designed to support independence over time. 
\end{abstract}




\section{Introduction}

Autistic adults face substantial barriers to employment and independent social participation, and these barriers are disproportionately social-communicative in origin. Studies consistently show that autistic individuals are underemployed relative to their cognitive abilities, and that workplace difficulties relate not to technical competence but to the social demands of the environment: navigating unwritten interpersonal norms, managing conversation timing, reading implied emotional signals, and communicating appropriately across different registers and relationships \cite{holwerda2012predictors,hendricks2010employment}. These difficulties do not resolve at the transition to adulthood; if anything, they intensify as the social contexts of adult life become less structured and more ambiguous than those of the school environment.
 
Social robots have been proposed as a tool for social-cognitive training, and a substantial body of HRI research has explored this direction \cite{pennisi2016autism,scassellati2012robots}. The evidence, however, is largely concentrated in child-focused interventions. Recent work has begun to extend this to adults: robot-mediated job interview training \cite{kumazaki2022robot,yoshikawa2023online} and workplace interruption-handling programmes \cite{ramnauth2022istar} have demonstrated that contextually situated robot training is both feasible and valued by autistic adults. Yet these studies remain few, and they address a narrow 
set of scenarios. The broader social-cognitive demands of adult autistic life, including managing conversation in ambiguous relational contexts, navigating unwritten workplace norms, and coping with the emotional and cognitive load of social camouflaging \cite{hull2017putting}, have not yet been systematically addressed through robot-based training. Crucially, the voice of autistic adults in shaping what that training should look like remains largely absent from the design process.
 
A second gap concerns who drives the design of these systems. Robot-based social interventions have been developed predominantly by researchers and clinicians, with autistic people positioned as evaluation subjects rather than design contributors. This approach risks producing systems that are technically sophisticated but ecologically misaligned: optimised for outcomes that researchers can measure rather than for difficulties that autistic adults actually experience. Participatory co-design, in which members of the target population are involved as active contributors to the design process, has been increasingly recognised as both an ethical imperative and an epistemic necessity in autism research \cite{fletcher-watson2019making,nicolaidis2011participatory}: autistic people hold first-person expertise about their own social-cognitive experience that is not recoverable through researcher observation or proxy report. The present study embeds this principle at the level of the research team: one team member is autistic and contributed to the study design, data interpretation, and theoretical framing as a researcher with lived experience.
 
This paper addresses both gaps. We report a focus group and co-design session conducted with five autistic adults, exploring their perspectives on social robots as tools for social-cognition training. The session combined open discussion with structured co-design activities on a shared digital whiteboard, generating both verbal and visual data. Our central research question is: \textit{what do autistic adults identify as the conditions, constraints, and design requirements for a social robot that would be genuinely useful for social-cognition training?} The paper makes two contributions. First, it reports seven themes derived from the co-design session that constitute a set of design requirements grounded in first-person adult autistic experience, covering training targets, feedback design, privacy and contextual limits, preparation strategies, and robot configurability. Second, it demonstrates the analytic value of co-design with autistic adults as a method for surfacing requirements that diverge from current field assumptions in specific, actionable, and theoretically grounded ways.

\section{Related Work}
 
\subsection{Autism in adulthood: social cognition and its challenges}
 
 
Social-cognitive difficulty in autism is distributed across multiple processes rather than localised in a single deficit. These include challenges in pragmatic language use and conversational timing \cite{adams2002speech}, difficulty reading non-verbal and implied emotional signals \cite{harms2010facial}, limited access to one's own emotional states (alexithymia), which affects approximately 50--65\% of autistic adults \cite{kinnaird2019alexithymia,bird2013mixed}, and difficulty integrating these processes in real time under the demands of live social interaction. Alexithymia in particular has been proposed to underlie many of the socio-emotional processing differences observed in autism independently of autism per se \cite{bird2013mixed}, with direct implications for how training tools should be designed: interventions that depend on introspective self-report at intake or during sessions may be systematically inaccessible to a majority of the target population.
A relevant theoretical reframing 
is Milton's double empathy problem \cite{milton2012ontological}, which repositions social difficulty as a bidirectional misunderstanding between autistic and non-autistic people rather than a deficit located in the autistic person. This perspective has practical implications for training design: it implies that social skills curricula built entirely around neurotypical interaction norms may be teaching autistic people to mask rather than communicate, and that interaction with other autistic people constitutes a distinct and underexplored training target. Together with work on camouflaging, which documents the cognitive, emotional, and social costs autistic adults bear when suppressing or imitating behaviour to pass as neurotypical \cite{hull2017putting,cage2019understanding}, this provides a principled argument for training environments that are private, non-judgemental, and oriented toward genuine competence rather than masking neurotypical norms.

\subsection{Social robots in autism intervention}
 
Socially assistive robots (SARs) have been explored as an intervention tool for autism, with evidence of engagement, attention, and specific skill gains primarily in children \cite{pennisi2016autism,scassellati2012robots}. The robot's predictable, simplified, and non-judgemental interactional style may reduce the anxiety associated with human social encounter and thereby lower the threshold for social engagement \cite{dautenhahn2007socially}. Reviews of the field consistently find, however, that the majority of studies involve children, use expression recognition from posed stimuli as the primary outcome measure, and report limited generalisation of gains to naturalistic settings when the robot is withdrawn \cite{pennisi2016autism,2024effectiveness}. The adult autistic population remains substantially underserved by this literature.
 
Recent work has begun to address more contextually grounded targets. Ramnauth et al.\ \cite{ramnauth2022istar} demonstrated that a social robot targeting workplace-relevant interruption-handling skills in adults with ASD produced measurable gains that users valued for their employment relevance. Kumazaki and colleagues \cite{kumazaki2022robot,yoshikawa2023online} developed and evaluated robot-mediated job interview training programmes for autistic individuals using android 
robots, demonstrating reductions in interview anxiety and improvements in performance when training targets were situated in realistic scenarios. Together these studies indicate that robot-based training can be effective in autistic adults when it moves beyond component-skill paradigms toward ecologically valid social scenarios.
Acceptance of SARs among autistic adults and relevant stakeholders is not unconditional. \ Frenkel et al. \cite{frenkel2025acceptance} underline the importance of centring autistic perspectives in deployment design, noting 
that autistic adults expressed more concerns about the robot-assisted scenario than other stakeholder groups, even though overall acceptance ratings did not significantly differ by group. This nuanced pattern underlines the importance of engaging autistic adults not only as end-users but as design evaluators throughout the development process.

Beyond intervention efficacy, recent HRI work has demonstrated social dynamics that apply directly to training contexts. Previous research has shown that people tend to disclose more to social robots over repeated sessions, accompanied by improvements in mood and an increasing perception of the robot as social and competent \cite{laban2024building,Laban2025CopingCaregivers}. Complementary findings indicate that negative emotional states can promote greater self-disclosure toward robots, suggesting that their non-judgemental consistency may be particularly valuable in the very emotional conditions most relevant to social-cognition training \cite{laban2023openingup}. Past research shows that repeated interaction with an LLM-powered social robot can guide emotional reflection and cognitive reappraisal, improving emotional understanding, perceived control, mood, and emotional expressiveness among students \cite{Laban2026}. Related work further frames social robots as supportive mediators for disabled students, including autistic students, particularly for signposting and disclosure-related interaction. In a within-subjects study, embodiment shaped perceived understanding, sociability, animacy, and privacy, while social effort and privacy remained key constraints on adoption \cite{Markelius2026SocialInteraction}. This body of evidence provides empirical grounding for the argument that the robot can function as a productive private rehearsal partner for populations who face barriers to disclosure in human interaction \cite{Laban2026a}. 
 
\subsection{Social skills training and feedback in autism}
 
Structured social skills training for autistic adults has a well-established evidence base, showing 
gains in social knowledge and social engagement \cite{laugeson2014peers,white2007social}. A persistent challenge is the generalisation of training gains from structured settings to naturalistic contexts \cite{cappadocia2012social}: a recent transfer analysis across 52 intervention studies found that ecological validity of the training context and the complexity of the targeted skills were among the strongest predictors of transfer \cite{2024effectiveness}. This finding motivates a shift away from decontextualised component-skill practice toward role-play-based approaches that simulate the pragmatic and contextual demands of real social encounters \cite{gaus2011cognitive,reichow2012social}. 
 
 
Participatory and co-design approaches have been increasingly recognised as epistemically necessary in autism research, not merely as ethical best practice \cite{fletcher-watson2019making,nicolaidis2011participatory}. Autistic adults bring first-person expertise about their own cognitive and social experience that is not recoverable through researcher observation or proxy report, and co-design sessions with this population have demonstrated capacity to generate technically specific, clinically grounded, and practically actionable design ~\cite{Ashburner2023Co-DesignAdults,Williams2023CyborgFunction}. The present study operationalises this principle through a structured focus group and co-design session with autistic adults, intended to inform the future design of social-robot social-cognition training.

\section{Method}
 
\subsection{Design}
 
We conducted an online focus group and co-design session via Zoom, lasting approximately 90 minutes. The session combined open discussion with structured co-design activities, following an agenda developed collaboratively by the research team that included an autistic researcher with lived experience (MB). The session was conducted in Hebrew and facilitated by a researcher in industrial engineering and management with experience in participatory design methods and human factors engineering. Ethical approval was obtained from the Ben-Gurion University of the Negev Institutional Ethics Board prior to data collection. The session was audio- and video-recorded for research purposes only, transcribed for analysis, and all data were anonymised prior to analysis and no identifying information is reported.
 
\subsection{Participants}
 
Five autistic adults participated (age range 18--52; five males). All participants had a formal diagnosis of autism spectrum disorder. Participants were recruited through an existing community network of autistic adults and provided informed consent before the session; consent was also verbally reviewed at the start of the online meeting. No compensation was provided and participation was voluntary. To protect anonymity, participants are referred to by speaker number throughout. One participant with higher support needs communicated in text during the session due to communication difficulties using the platform, and was supported by his parent. Their contributions were read aloud by the parent and the facilitator and treated equivalently to verbal contributions. The sampling composition should be considered when interpreting the scope of the findings. Male predominance in diagnosed autism samples has been discussed as requiring etiological explanation \cite{baroncohen2011why}; however, diagnosed and community-recruited samples may also underrepresent autistic women and girls because of later recognition, camouflaging, and diagnostic bias \cite{mccrossin2022finding}. Future co-design work should therefore include autistic women and gender-diverse adults.
 
\subsection{Session procedure}
 
The session comprised four sequential phases. The first phase was an icebreaker activity in which participants were invited to describe an imaginary ``\textit{activation button}'' they would want added to their lives, designed to elicit reflection on social-cognitive challenges and establish a safe conversational environment. The second phase was an open focus group discussion on social robots: initial associations, desired use contexts, situations where a robot should not be present, robot design preferences, and preparation strategies for social encounters. The third phase was a structured co-design activity conducted on a shared Miro digital whiteboard. Participants contributed virtual sticky notes across four boards: desired feedback content; preferred feedback format; preferred timing of robot intervention; and preferred robotic expression. The final phase was an open closing discussion in which participants could share additional ideas, reflections, or concerns beyond the structured activities. 
 
 
 
\subsection{Analysis}
 
Transcripts and Miro board content were analysed using reflexive thematic analysis \cite{braun2006thematic}. Three researchers conducted open coding of the transcript, generating initial codes inductively from the data without a predetermined coding frame. The research team included an autistic researcher, whose lived-experience expertise informed some of the interpretation of the data. Codes were grouped into candidate themes, which were then reviewed against the full dataset, refined, and named. Miro board outputs were analysed as a complementary data source: the categories and content of sticky notes were coded and integrated into the thematic structure where they corroborated, extended, or nuanced the verbal data. Analysis was conducted in Hebrew and themes were translated into English for reporting; all participant quotations were translated and checked for accuracy. 

\begin{figure*}[h!]
    \centering
    \includegraphics[width=\linewidth]{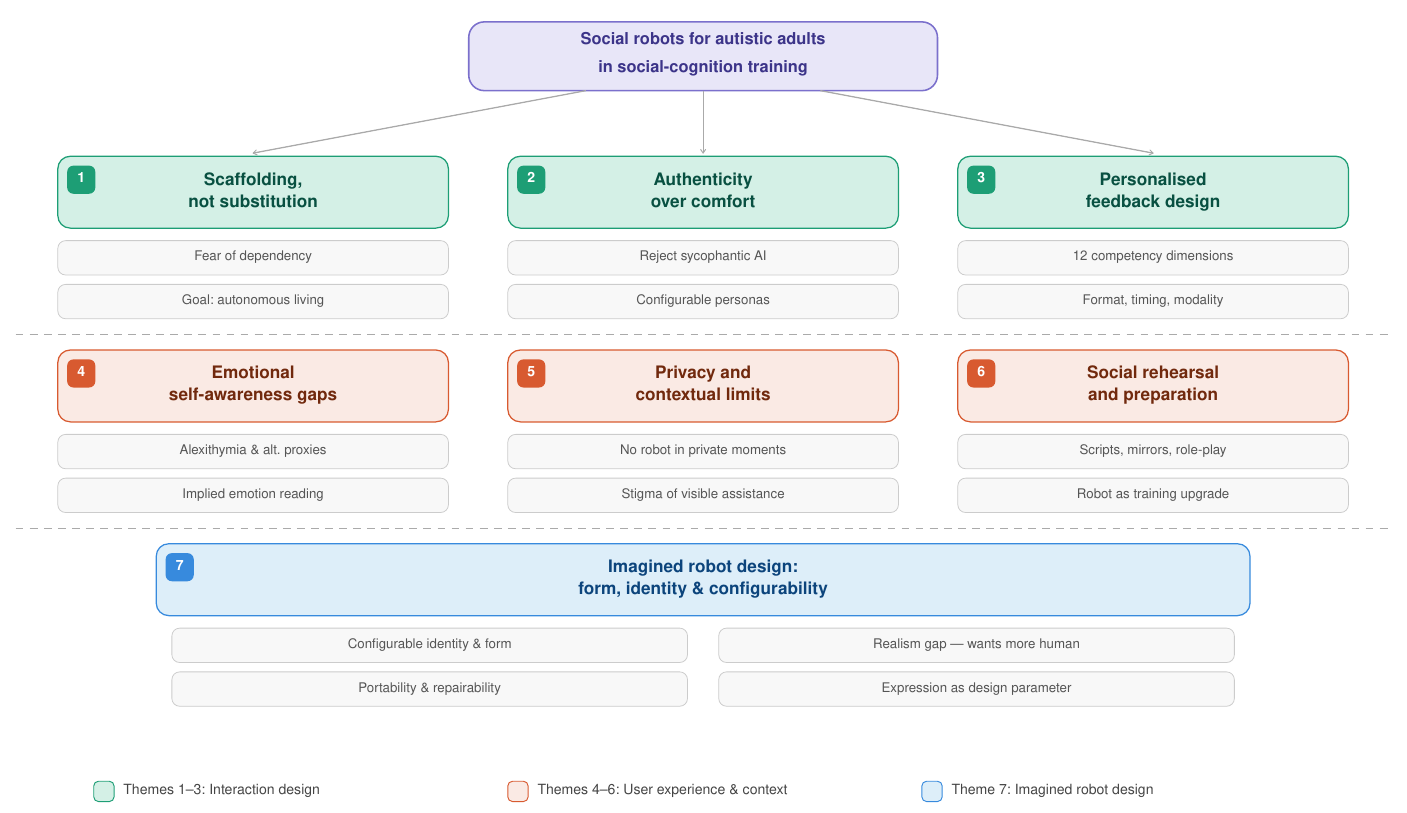}
    \caption{Thematic map of the seven themes identified in the analysis.}
    \label{fig:map}
\end{figure*}

\section{Results}

\begin{figure*}[!t]
    \centering
    \includegraphics[width=\linewidth]{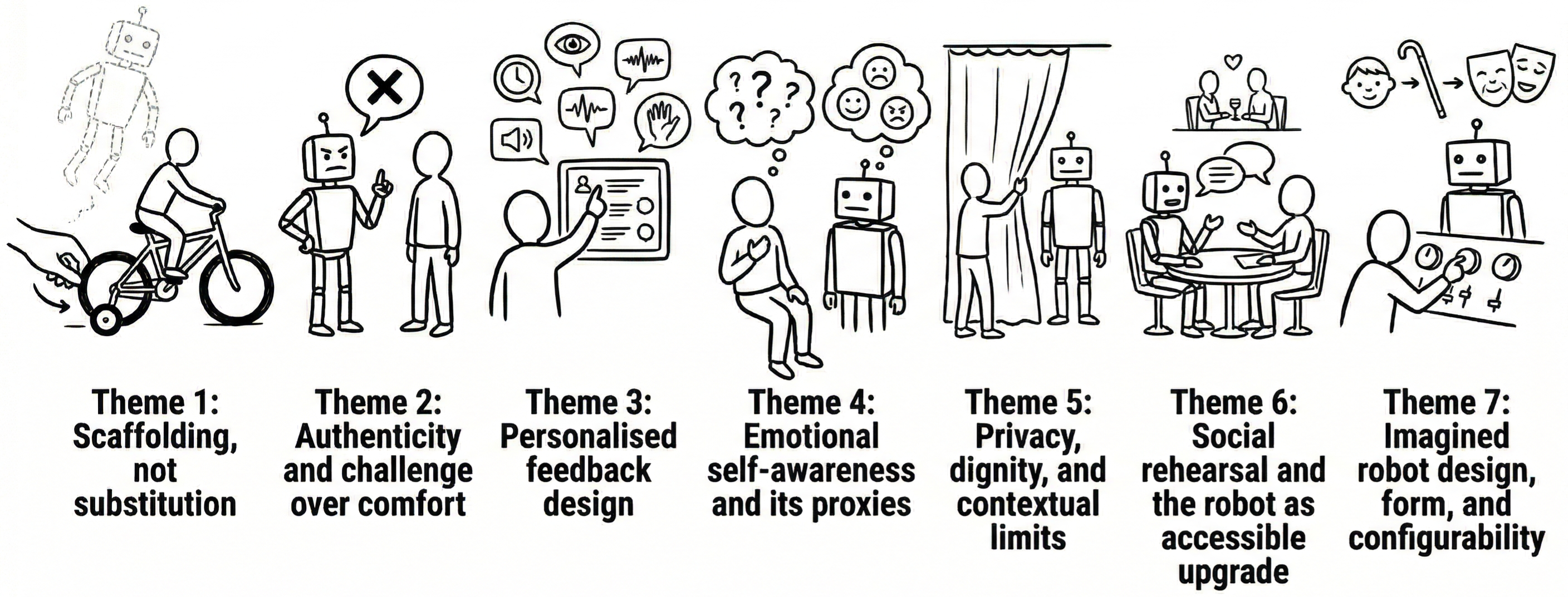}
    \caption{\footnotesize Visual summary of the seven themes generated through the focus group and co-design session with autistic adults, illustrating the core design requirements for a social robot for social-cognition training: scaffolding rather than substitution, authenticity over comfort, personalised feedback, support for emotional self-awareness gaps, privacy and contextual limits, social rehearsal, and configurability in robot identity, form, and expression.}
    \label{fig:placeholder}
\end{figure*}

\begin{figure*}[h!]
    \centering
  \includegraphics[width=\linewidth]{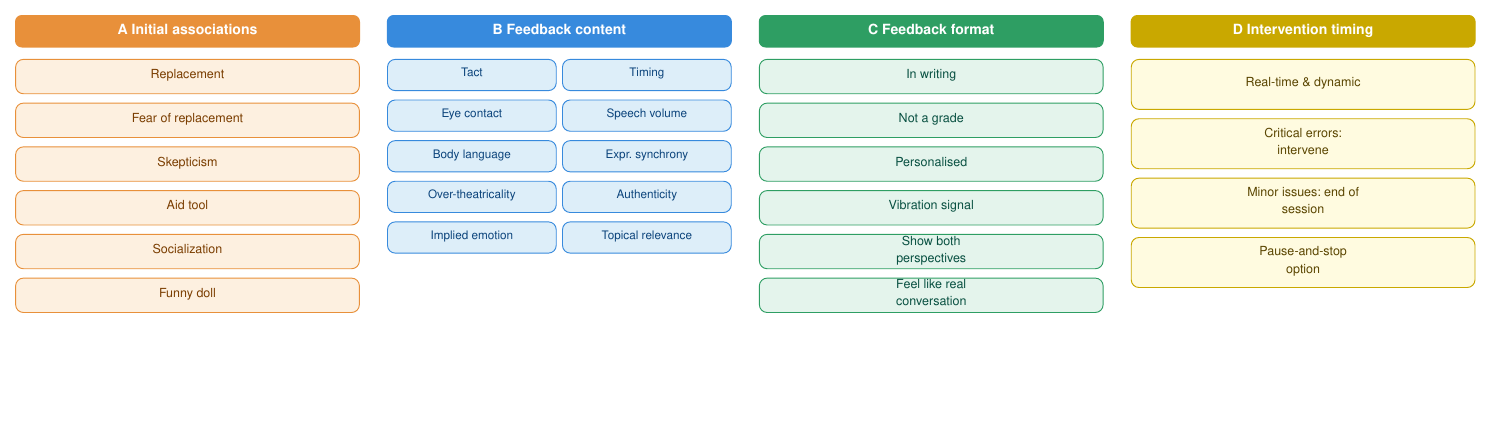}
    \caption{Co-design board outputs from the Miro  whiteboard activity. Panel~A shows word associations with ``social robot.'' Panel~B lists communication dimensions participants requested feedback on. Panel~C shows preferred feedback format. Panel~D shows preferred intervention timing rules.}
    \label{fig:miro}
\end{figure*}

Thematic analysis yielded seven themes organised across two tiers (see Figure~\ref{fig:map}). 
The first tier (Themes~1--3) concerns the robot's role and interaction design. The second tier (Themes~4--6) concerns participants' lived experience of social-cognitive difficulty and existing coping strategies. Theme~7 addresses how participants imagined the robot as a designed artefact and is shared across the two tiers. See Figure~\ref{fig:placeholder} for an illustration of the themes, and Table~\ref{tab:themes} for their primary design implications. Throughout, participants are referred to by speaker number (P1, P2, P4, P5, P6). All quotations are translated from Hebrew. Miro board outputs are referenced where they corroborate verbal contributions (see Figure~\ref{fig:miro}). Because the data were generated in a group setting, we also note how some ideas developed through sequential elaboration. For example, the concern that robots could replace human support was first raised as a negative association, then reframed as a training-tool requirement, and later elaborated into the principle that the robot should support independence rather than dependency. Similar patterns appeared around privacy boundaries and non-sycophantic feedback, while feedback format showed explicit variation across participants.

\subsection{Theme~1: Scaffolding, not substitution}

The word \textit{replacement} surfaced before any design properties had been discussed. P1: \textit{``The connotation for me is not the most positive, more as a replacement. It scares me a little to replace people with robots.''} The Miro word-association board confirmed this was not isolated: prominent initial associations included \textit{replacement}, \textit{fear of replacement}, and \textit{skepticism}, alongside \textit{aid tool} and \textit{socialization}. P2 reframed it immediately: \textit{``I would more want the robot not to be a replacement, but a training tool.''} P1 articulated it through analogy: \textit{``It's like when a child starts learning to ride a bike, only at the beginning he has four wheels. After time he needs to learn to ride on two. But if he always rides with four wheels... with AI it becomes a tool that then becomes a substitute.''} P4 named the underlying mechanism: \textit{``The more you have tools that do things for you, the more people generally get lazy. A good example was AI through Google} [i.e., Large Language Models in web or mobile interfaces, like ChatGPT or Gemini]\textit{: people lose vital abilities.''} The resolution was unambiguous: the robot should be designed to self-obsolete, beginning with heavy scaffolding and progressively withdrawing support as the user gains competence. P2: \textit{``The robot should teach us to live without it.''} This framing aligns with fading-prompt approaches in social skills intervention \cite{laugeson2014peers,white2007social} and with self-determination theory's (SDT) emphasis on autonomous motivation over extrinsic dependency \cite{ryan2000self}.

\subsection{Theme~2: Authenticity and challenge over comfort}

Participants criticised the agreeable tone of current AI systems as a training liability. P4: \textit{``The AI always agrees with you, always apologises. The brain gets lazy. You learn that you succeed more than you'll actually succeed with other people.''} P1 was equally direct: \textit{``A real person will say their opinion how they want. They might even curse. They'll have opposition. They don't like what you're saying. That's what a robot can't exactly express.''} What participants were describing is a concern about ecological validity in training stimuli. A socially unrealistic training partner, one that never disagrees, never misreads, and never shows impatience, produces social skills calibrated to an interaction partner that does not exist. P6 independently named the same concern: \textit{``The robot should also have some opposition --- not always agree with everything.''} He also noted that he currently uses AI tools only for technical tasks, not for social interaction, addressing the sycophancy problem as the reason \cite{cheng26,Teixeira2026}. The proposed solution was a robot with rotating, user-configurable personas spanning age, gender, cultural background, and formality register. P1: \textit{``If you want a formal setting, like preparing to talk with a government official, you set it to formal. That way it's preparing you for that thing.''} P2 extended this: \textit{``All kinds of traits, for example a Polish man aged 80 from Ramat Gan} [a city in Israel]\textit{, and then you practice, so you become a kind of adapter to every situation.''}  Participants also requested explicit assertiveness training through robot-initiated disagreement.  P2 added: \textit{``The robot should know how to hold positions, and we can argue with it, and through that we can learn skills of how to stand up for yourself without hurting the other side. That's called assertiveness.''} A goal-declaration mechanism was proposed: before each session, the user states the intended social outcome (for example, \textit{``I want the interviewer to see me as reliable''} or \textit{``I want her to know I'm interested romantically''}) so feedback can be evaluated against it rather than a generic standard. This addresses documented difficulties in autism with social intention-reading \cite{baron-cohen1995mindblindness} and connects to the ecological validity problem that has limited generalisation in structured social skills training \cite{cappadocia2012social}.

\subsection{Theme~3: Personalised feedback design}

The Miro co-design board generated twelve communication dimensions participants wished to receive feedback on, including tact, speech timing, eye contact, topical relevance, body language reading, verbal and non-verbal synchrony, over-theatricality, and the ability to read implied emotional states. This breadth is significant: it positions social-cognitive difficulty as distributed across pragmatics, self-monitoring, and contextual appropriateness rather than localised in facial expression recognition, which has dominated the robot-autism literature \cite{pennisi2016autism}. On feedback format, the green Miro cards showed strong preference for personalised, non-numeric delivery; P4 rejected scoring outright (\textit{``I don't think behaviour like ours can be measured on a yes/no scale''}), while P2 preferred numerical output, confirming the need for user-selectable format. On timing, the yellow cards captured a severity-threshold rule: critical errors warrant real-time interruption; minor issues should accumulate for end-of-session review. This maps onto evidence that terminal feedback produces better long-term retention by preserving internal error-detection processes \cite{schmidt1990frequent}, while acknowledging that the review window may be too wide for populations with reduced working memory capacity for social cues \cite{Barendse2013}. A revealing contrast emerged when participants were asked whether they would give a friend feedback in the same way they wanted it from the robot. All said no. P4: \textit{``I would start in a completely different way... I adapt myself to the environment.''} P1 cited the fear of causing hurt: \textit{``Unlike the robot, which has no fears — it can be objective without feelings.''} P2 explained that for him, a robot inhabits a different register from friendship precisely because it does not need to comply with the social codes of social relationships. The implication is that the robot's value lies partly in its \textit{non-humanness}: it can deliver the honest, direct feedback that human social relationships structurally cannot. Interestingly, P6  specified written text as his preferred mode of receiving feedback.

\subsection{Theme~4: Emotional self-awareness and its proxies}

P2 introduced alexithymia by name, saying: \textit{``
we don't always have access to our own emotional states.''} Alexithymia, characterised by difficulty identifying and describing subjective feelings, affects an estimated 50--65\% of autistic adults and has been proposed to underlie many socio-emotional processing differences in the population independently of autism per se \cite{kinnaird2019alexithymia,bird2013mixed}. The concept arrived in the session through lived experience before it was named. P4 described wishing for \textit{``a button that would allow me to put my feelings into words, there are many times I feel something and simply don't know how to explain it.''} He also described a second button: \textit{``Something that when I press it the sensation becomes more pleasant. There's a kind of discomfort, or something's not right, or a bit of a mess in my head. If there were a button that would lower that, that could be nice.''} Following P4, participants described the resulting social difficulty in relational terms: P2 could not detect when his wife was upset unless she stated it explicitly; P1 could not determine whether a text message was flirtatious or merely friendly. These examples point to a training target that is poorly served by existing technology: the low-intensity, ambiguous, contextually embedded emotional signals of everyday close relationships rather than the posed, high-intensity expressions used in most evaluation paradigms \cite{harms2010facial,golan2006cambridge}. 
This makes such signals especially relevant for adult social functioning in close relationships and professional settings.
The theme's most creative design contribution addressed the alexithymia-specific self-report bottleneck directly: P6 proposed that the robot should infer the user's current emotional state from their music listening behaviour, connecting to Spotify or YouTube Music, without requiring introspective reporting. This extends affective computing's shift toward behavioural proxies for emotional state in a user-directed, consent-based direction \cite{Onnela2016HarnessingHealth}.

\subsection{Theme~5: Privacy, dignity, and contextual limits}

Unlike the other themes, which involved negotiation and design preference, the question of where a robot should \textit{not} be present produced immediate, categorical responses. P1 answered before the question was complete: \textit{``In the middle of a date, at least if it's physical.''} P4 confirmed: \textit{``Private moments, yes.''} P1 elaborated: \textit{``In the middle of doing a private action, let's tell the truth. Nobody wants to be looked at in such a private situation.''} When the facilitator redirected the question to assume a fully secure, non-recording robot, participants maintained their boundaries, confirming that the exclusion zones are grounded in dignity and relational appropriateness rather than in distrust of technology. P1 offered a generalised heuristic: \textit{``Any situation where you wouldn't take your mother, there's something in that.''} This heuristic sidesteps the enumeration problem by operationalising the exclusion criterion as the presence of relational intimacy or privacy that would be violated by any third-party observer, human or robotic. P4 identified a subtler concern: \textit{``What impression does it make if you have a little robot whispering things in your ear?''} He invoked the image of Jiminy Cricket coaching Pinocchio to make the point vivid: visible assistive technology does not simply signal disability; it signals a dependency on external coaching to perform basic social interaction, which carries its own stigma. P1 extended this into a social contagion argument: \textit{``People might also suspect people who are using this even when they're not.''} This is a stigma externality: the visibility of an assistive device can generate heightened scrutiny of others who resemble its typical users. This connects to Goffman's account of stigma management \cite{goffman1963stigma} 
as P2 articulated the design implication through Goffman's front stage and back stage framework \cite{goffman1959presentation}: the robot belongs in the rehearsal space, not the performance. This should be treated as a deployment constraint, not a preference.

\subsection{Theme~6: Social rehearsal and the robot as accessible upgrade}

Participants described sophisticated but resource-intensive self-developed preparation strategies: role-play with a parent who is a therapist (P1: \textit{``I know it's a hell of a privilege to have a therapist at home''}), pre-reading scientific literature on social interaction (P2), and deliberate self-presentation to interviewers (P4). These strategies map directly onto the components of structured social skills programmes \cite{laugeson2014peers,gaus2011cognitive}, but depend on privileged access. P2 described a recurring failure mode: preparation that broke down not from lack of knowledge but from misreading the social temperature in real time, leading to contextually inappropriate statements in workplace settings. This illustrates a dissociation between declarative social knowledge and procedural social competence that behavioural training targets through situated practice \cite{reichow2012social}. Participants were therefore not asking for instruction in social rules they do not know; they were asking for a tool that would help them bridge the gap between knowledge and situated performance. P4 closed this part of the session by proposing a contextual extension with significant practical implications. He had attempted military service but found the social demands overwhelming: \textit{``The content, I knew and everything, but the social demands, standing at attention, the timing of group interaction, the others, that was really hard and didn't suit me well. But if [the robot] could help with things like that...''} Therefore, the robot was framed not as a foreign intervention but as an accessible, on-demand upgrade to practices already validated through experience, extending naturally to workplace contexts where social and communication demands are a primary barrier to sustained employment for autistic adults \cite{holwerda2012predictors,hendricks2010employment}.

\subsection{Theme~7: Imagined robot design}

Before seeing any specific robot, participants converged on a shared design vision: the robot as a configurable platform whose interactional identity is under user control rather than fixed. P2: \textit{``I would want the robot to know how to adjust its gender, age, and appearance depending on the situation.''} P1 extended this to cultural specificity: \textit{``In the same way it can be anyone you choose.''} This vision is consistent with arguments in HRI that social robots should function as interaction platforms rather than agents with predetermined personalities \cite{dautenhahn2007socially,bransky_identity_2026}, and reflects the insight that social skill is context-sensitive adaptation, not a unitary competence \cite{adams2002speech}. Two further design properties were raised: portability (a pocket-sized form factor for use outside structured sessions) and repairability, with P5 requesting open-source maintainability to address the economic barriers autistic adults disproportionately face \cite{austin2017neurodiversity}. Expression controllability emerged as a third property. Participants proposed that the legibility of the robot's expressive behaviour, spanning face, voice, gesture, and their synchrony, should be a user-adjustable training parameter. 
The design implication is an adaptive system in which expression parameters are set at intake and adjusted as evidence of learning accumulates, a direction we will explore empirically in future work.

\begin{table}[t]
\caption{Summary of seven themes and primary design implications}
\label{tab:themes}
\renewcommand{\arraystretch}{1.35}
\begin{tabular}{@{}p{0.28\columnwidth}p{0.64\columnwidth}@{}}
\toprule
\textbf{Theme} & \textbf{Primary design implication} \\
\midrule
1.\ Scaffolding, not substitution &
  Design for progressive independence; measure success by transfer to unaided performance, not engagement with the robot. \\
2.\ Authenticity over comfort &
  The robot should disagree, hold positions, and resist sycophancy; its non-humanness licenses honest feedback that human relationships cannot sustain. \\
3.\ Personalised feedback design &
  Target distributed pragmatic competencies (tact, timing, body language, synchrony); allow user-selectable format and severity-threshold timing. \\
4.\ Emotional self-awareness gaps &
  Route around alexithymia by inferring emotional state from behavioural proxies rather than requiring introspective self-report. \\
5.\ Privacy and contextual limits &
  Confine the robot to private preparatory sessions; visible in-situ deployment would compound the camouflaging burden autistic adults already carry. \\
6.\ Social rehearsal and preparation &
  The robot is an accessible, on-demand upgrade to existing self-developed preparation strategies; ecological validity of scenarios is critical. \\
7.\ Imagined robot design &
  Support configurable interactional identity (age, gender, register); treat expression legibility as a user-adjustable training parameter. \\
\bottomrule
\end{tabular}
\end{table}



\section{Discussion and Conclusion}

The findings present a coherent design vision from the target population, and one that sits in productive tension with the current state of the field. The design vision participants articulated diverges in notable ways from the framings that dominate existing robot-autism work. Two divergences were explicit: they envisioned a time-limited scaffold rather than a persistent companion \cite{pennisi2016autism,scassellati2012robots}, and they described needs anchored in adult social contexts that the predominantly child-focused literature does not address \cite{pennisi2016autism}. A third divergence is analytical: the competencies they described as most important, such as tact, conversational timing, implied emotion reading, and cross-channel expressive synchrony, are not the competencies that robot-autism research has predominantly targeted \cite{pennisi2016autism}. Across these themes, participants consistently described a robot that serves as a private rehearsal partner for situated social scenarios, and that is explicitly designed to become redundant as the user gains independence \cite{ryan2000self}. The significance of this is not that participants identified user preferences, but that they articulated a fundamentally different theory of what the training is for. The themes also suggest that the robot should not be understood as occupying a single role. Participants imagined a bounded set of roles: a backstage rehearsal partner, a configurable role-play interlocutor, a direct feedback provider, and a fading scaffold. These roles impose embodiment constraints: the robot should be available for private rehearsal and configurable in identity and expression, but largely absent from live front-stage social situations where users want to perform independently.

The training-target profile described by participants, including tact, conversational timing, implied emotion reading, and cross-channel expressive synchrony, maps onto distributed social-cognitive difficulties documented in the autism literature \cite{adams2002speech,klin2003social,mcpartland2011social}, while moving away from the component-skill paradigms that transfer analysis has shown to generalise least well \cite{2024effectiveness}. Recent adult-focused HRI programmes show that contextually grounded robot training is feasible \cite{ramnauth2022istar,kumazaki2022robot,yoshikawa2023online}; the present findings add a first-person account of what such targets should look like in practice.

 
A related finding concerns the robot's authenticity as a feature rather than a limitation. Participants wanted the robot to disagree, hold positions, and resist the excessive agreeableness they identified in current AI systems. Yet when asked whether they would give a friend the same kind of honest, direct feedback they wanted from the robot, all said no. 
For autistic adults, the robot's \textit{non-humanness} is precisely what licenses it to be authentic in ways that human social relationships structurally cannot sustain. This is not a workaround for an imperfect substitute, it is a genuine affordance. Human feedback is always mediated by the social costs of honesty, the fear of causing hurt, and the codes that govern front-stage interaction \cite{goffman1959presentation}. The robot, operating backstage, is exempt from those costs. Designing it to be comforting and agreeable would squander that affordance.
 
The alexithymia finding extends the training-target argument in a direction that HRI has not yet fully addressed. If a substantial proportion of autistic adults cannot reliably self-report their emotional state, then training systems that open with ``how are you feeling?'' and require users to track and verbalise internal experience during practice are structurally inaccessible to a significant part of the population. As previous findings on self-disclosure \cite{laban2024building,laban2023openingup,Laban2026} suggest that the robot's non-judgemental consistency is precisely the condition that allows disclosure to emerge over time without requiring it up front, future interventions should utilise it. The participant proposal to infer emotional state from behavioural proxies such as music listening behaviour is a design-level instantiation of the same insight: whether through music or other consent-based behavioural channels, the broader implication is to route around the self-report bottleneck rather than demand introspective access.
 
The anti-dependency concern raised independently by multiple participants is not scepticism about technology; it is a precise articulation of a documented failure mode: generalisation of robot-mediated gains beyond clinical environments remains the field's central unresolved problem \cite{kewalramani2023}, and self-determination theory explains why: when the robot substitutes for internal regulation, the autonomy the intervention is meant to build does not develop \cite{ryan2000self}. Participants were, in effect, asking the field to measure the right thing: not engagement during training, but independent performance after it. The privacy and stigma boundaries participants drew are 
deployment constraints that should be considered. 
Autistic adults already bear the cognitive and emotional costs of camouflaging in live social encounters \cite{hull2017putting,cage2019understanding}; adding a visible assistive device to those encounters would compound rather than reduce that burden. 
As P2 said, the robot belongs in the back-stage, 
and measuring its success requires looking at what happens on the front stage without it. 

These findings also clarify how the proposed design agenda relates to the current state of AI for robotics. Some aspects of the envisioned system are feasible with current technology if framed as a constrained rehearsal environment rather than as a fully autonomous social coach: the robot can support private role-play, adjustable interaction styles, user-declared goals, staged scenarios, and gradual reduction of assistance. However, the more consequential design features depend on forms of social and affective inference that remain technically fragile 
in open-ended interactions. These should therefore be treated not as assumed capabilities, but as targets for human-supervised design and empirical validation.

The seven themes together constitute a set of design requirements that are specific, grounded in first-person social-cognitive experience, and partly addressable by current robot platforms when framed as constrained rehearsal systems. The clearest practical implication is that the training target needs to shift: from component skills toward situated pragmatic scenarios, and from persistent monitoring toward a scaffold that measures its own success by the pace at which it becomes unnecessary. Taken together, the findings establish what autistic adults themselves identify as the conditions under which a SAR would be genuinely useful, providing a first-person grounded basis for adult-focused robot design.

\section*{Acknowledgment}
This research was supported by Ben-Gurion University of the Negev through the Agricultural, Biological and Cognitive Robotics Initiative (funded by the Marcus Endowment Fund and the Helmsley Charitable Trust).

\balance
\bibliographystyle{IEEEtran}
\bibliography{social_robot_roman}

\end{document}